\documentstyle[12pt]{article}
\begin{document}
\tolerance=5000
\def\pp{{\, \mid \hskip -1.5mm =}}
\def\cL{{\cal L}}
\def\be{\begin{equation}}
\def\ee{\end{equation}}
\def\bea{\begin{eqnarray}}
\def\eea{\end{eqnarray}}
\def\tr{{\rm tr}\, }
\def\nn{\nonumber \\}
\def\e{{\rm e}}
\def\D{{D \hskip -3mm /\,}}

\def\SEH{S_{\rm EH}}
\def\SGH{S_{\rm GH}}
\def\AdS5{{{\rm AdS}_5}}
\def\S4{{{\rm S}_4}}
\def\gfv{{g_{(5)}}}
\def\gfr{{g_{(4)}}}
\def\SC{{S_{\rm C}}}
\def\RH{{R_{\rm H}}}

\def\wlBox{\mbox{
\raisebox{0.3cm}{$\widetilde{\mbox{\raisebox{-0.3cm}{$\Box$}}}$}}}

\  \hfill 
\begin{minipage}{3.5cm}
%NDA-FP-?? \\
September 2000 \\
%hep-th/yymmxxx \\
\end{minipage}

\vfill

\begin{center}
{\large\bf 
Dilatonic quantum multi-brane-worlds.
}

\vfill

{\sc Shin'ichi NOJIRI}\footnote{email: nojiri@cc.nda.ac.jp}, 
 {\sc Sergei D. ODINTSOV}$^{\spadesuit,\clubsuit}$\footnote{
email: odintsov@ifug5.ugto.mx, odintsov@itp.uni-leipzig.de}, \\
and {\sc K.E. OSETRIN$^{\spadesuit}$}\footnote{
e-mail: osetrin@tspu.edu.ru}

\vfill

{\sl Department of Applied Physics \\
National Defence Academy, 
Hashirimizu Yokosuka 239, JAPAN}

\vfill

{\sl $\spadesuit$ 
Tomsk State Pedagogical University,
634041 Tomsk,RUSSIA}

\vfill

{\sl $\clubsuit$ Instituto de Fisica de la Universidad de 
Guanajuato \\
Apdo.Postal E-143, 37150 Leon, Gto., MEXICO} 

\vfill

{\bf ABSTRACT}

\end{center}

d5 dilatonic gravity action with surface counterterms motivated by 
AdS/CFT correspondence and with contributions of brane quantum CFTs is 
considered around AdS-like bulk. The effective equations of motion are 
constructed. They admit two (outer and inner) or multi-brane solutions 
where brane CFTs may be different. The role of quantum brane CFT is in 
inducing of complicated brane dilatonic gravity. For exponential bulk 
potentials the number of AdS-like bulk spaces is found in analytical form.
The correspondent flat or curved (de Sitter or hyperbolic) dilatonic 
two branes are created, as a rule, thanks to quantum effects. The 
observable early Universe may correspond to inflationary brane. The found 
dilatonic quantum two brane-worlds usually contain the naked singularity 
but in couple explicit examples the curvature is finite and horizon 
(corresponding to wormhole-like space) appears.

\vfill 

\noindent
PACS number: 04.50.+h; 04.70.-s; 98.80.Cq; 12.10.Kt

\newpage

\section{Introduction}

Recent booming activity in the study of brane-worlds is caused 
by several reasons. First, gravity on 4d brane embedded in 
higher dimensional AdS-like Universe may be localized \cite{RS1,RS2}. 
Second, the way to resolve the mass hierarchy problem 
appears\cite{RS1}. Third, the new ideas on 
cosmological constant problem solution come to game \cite{ADKS,ADDK}.
Very incomplete list of references\cite{CH,cosm} (and references therein) 
mainly on the cosmological aspects of brane-worlds is growing 
every day.

The essential element of brane-world models is the presence 
in the theory of two free parameters (bulk cosmological constant 
and brane tension, or brane cosmological constant). The role of 
brane cosmological constant is to fix the position of the brane 
in terms of tension (that is why brane cosmological constant 
and brane tension are almost the same thing). 
Being completely consistent and mathematically reasonable, 
such way of doing things may look not completely satisfactory. 
Indeed, the physical 
origin (and prediction) of brane tension in terms of some dynamical 
mechanism may be required.

The ideology may be different, in the spirit of refs.\cite{NOZ, HHR}.
One considers the addition of surface counterterms to the 
original action on AdS-like space. These terms are responcible 
for making the variational  
procedure to be well-defined (in Gibbons-Hawking spirit) and for 
elimination of the leading divergences of the action. 
Brane tension is 
not considered as free parameter anymore but it is fixed by the 
condition of finiteness of spacetime when brane goes to 
infinity. Of course, leaving 
the theory in such form would rule out the possibility of consistent 
brane-world solutions existance. Fortunately, 
other parameters contribute  
to brane tension. If one considers that there is quantum 
CFT living on 
the brane (which is more close to the spirit of AdS/CFT 
correspondence\cite{AdS} ) then such CFT produces conformal anomaly 
(or anomaly induced effective action). This contributes to 
brane tension.
As a result dynamical mechanism to get brane-world with flat or 
curved (de Sitter or Anti-de Sitter) brane appears. 
The curvature of such 4d Universe is 
expressed in terms of some dimensional parameter $l$ which 
usually appears in AdS/CFT set-up and of content of quantum 
brane matter. In other words,
brane-world is the consequence of the fact (verified 
experimentally by everybody life) of the presence of 
matter on the brane! For example, sign of conformal anomaly 
terms for usual matter is such that in 
one-brane case the de Sitter (ever expanding, inflationary ) 
Universe is preferrable solution of brane 
equation\footnote{Similar mechanism for 
anomaly driven inflation in usual 4d world has been invented by 
Starobinsky\cite{SMM} and generalized for dilaton presence in 
refs.\cite{Brevik}}. 

The scenario of refs.\cite{NOZ,HHR} may be extended to the presence
of dilaton(s) as it was done in ref.\cite{NOO} or to formulation of 
quantum cosmology in Wheeler-De Witt form \cite{ANO}. 
Then whole scenario looks even more related with AdS/CFT 
correspondence as dilatonic gravity naturally follows as bosonic 
sector of d5 gauged supergravity. Moreover, the extra prize-in 
form of dynamical determination of 4d boundary value 
of dilaton-appears. In ref.\cite{NOO} the quantum dilatonic one brane 
Universe has been presented with possibility to get inflationary or 
hyperbolic or flat brane with dynamical determination of brane 
dilaton. The interesting question is related with 
generalization of such scenario in dilatonic gravity for 
multi-brane case. This will be the purpose of present work.

In the next section we present general action of d5 dilatonic gravity 
with surface counterterms and quantum brane CFT contribution. This action 
is convenient for description of brane-worlds where bulk is AdS-like 
spacetime. There could be one or two (flat or curved) branes in the theory.
As it was already mentioned the brane tension is fixed in our approach,
instead of it the effective brane tension is induced by quantum effects.
Section three is devoted to formulation of effective bulk-brane field 
equations. The explicit analytical solution of bulk equation for number 
of exponential bulk potentials is presented. The lengthy analysis of 4d 
brane equations shows the possibility to have two (inner and outer) 
branes associated with each of above bulk solutions. It is interesting 
that quantum created branes can be flat, or de Sitter (inflationary) or 
hyperbolic.
The role of quantum brane matter corrections in getting of such branes is
extremely important. Nevertheless, there are few particular cases where 
such branes appear on classical level, i.e. without quantum corrections.
In section four we briefly describe how to get generalization of above 
solutions for quantum dilatonic multi-brane-worlds
with more than two branes. Brief summary of results is given in final 
section where also the study of character of singularities for proposed 
two-brane solutions is presented. In most cases, as usually occurs in AdS 
dilatonic gravity, the solutions contain the naked singularity. However, 
in couple cases the scalar curvature is finite and there is horizon.
The corresponding 4d branes may be interpreted as wormhole.

\section{Dilatonic gravity action with brane quantum corrections}

Let us present the initial action for dilatonic AdS gravity under 
consideration. The metric of (Euclidean) AdS has the following form:
\be
\label{AdS}
ds^2=dz^2 + \sum_{i,j=1}^4 g_{(4)ij} dx^i dx^j\ ,
\quad g_{(4)ij}=\e^{2\tilde A(z)}\hat g_{ij}\ .
\ee
Here $\hat g_{ij}$ is the metric of the Einstein manifold, which is
defined by $r_{ij}=k\hat g_{ij}$, where $r_{ij}$ is 
the Ricci tensor constructed with $\hat g_{ij}$ and 
$k$ is a constant. 
One can consider two copies of the regions given by $z<z_0$ and 
glue two regions  putting a brane at $z=z_0$. 
More generally, one can consider two copies of regions 
$\tilde z_0<z<z_0$ and glue the regions putting two branes 
at $z=\tilde z_0$ and $z=z_0$. 
Hereafter we call the brane at $z=\tilde z_0$ as ``inner'' brane 
and that at $z=z_0$ as ``outer'' brane. 

Let us first consider the case with only one brane at $z=z_0$ and 
start with Euclidean signature action $S$ which is 
the sum of the Einstein-Hilbert action $\SEH$ with kinetic term 
and potential $V(\phi)={12 \over l^2}+\Phi(\phi)$ for dilaton 
$\phi$, the Gibbons-Hawking surface term $\SGH$,  the surface 
counter term $S_1$ and the trace anomaly induced action 
$W$\footnote{For the introduction to anomaly induced 
effective action in curved space-time (with torsion), see
section 5.5 in \cite{BOS}. This anomaly induced action is due to 
brane CFT living on the boundary of dilatonic AdS-like space.}: 
\bea
\label{Stotal}
S&=&\SEH + \SGH + 2 S_1 + W, \\
\label{SEHi}
\SEH&=&{1 \over 16\pi G}\int d^5 x \sqrt{\gfv}\left(R_{(5)} 
 -{1 \over 2}\nabla_\mu\phi\nabla^\mu \phi 
 + {12 \over l^2}+ \Phi(\phi) \right), \\
\label{GHi}
\SGH&=&{1 \over 8\pi G}\int d^4 x \sqrt{\gfr}\nabla_\mu n^\mu, \\
\label{S1}
S_1&=& -{1 \over 16\pi G l}\int d^4 x \sqrt{\gfr}\left(
{6 \over l} + {l \over 4}\Phi(\phi)\right), \\
\label{W}
W&=& b \int d^4x \sqrt{\widetilde g}\widetilde F A \nn
&& + b' \int d^4x\sqrt{\widetilde g}
\left\{A \left[2{\wlBox}^2 
+\widetilde R_{\mu\nu}\widetilde\nabla_\mu\widetilde\nabla_\nu 
 - {4 \over 3}\widetilde R \wlBox^2 
+ {2 \over 3}(\widetilde\nabla^\mu \widetilde R)\widetilde\nabla_\mu
\right]A \right. \nn
&& \left. + \left(\widetilde G - {2 \over 3}\wlBox \widetilde R
\right)A \right\} \\
&& -{1 \over 12}\left\{b''+ {2 \over 3}(b + b')\right\}
\int d^4x \sqrt{\widetilde g} 
\left[ \widetilde R - 6\wlBox A 
 - 6 (\widetilde\nabla_\mu A)(\widetilde \nabla^\mu A)
\right]^2 \nn
&& + C \int d^4x \sqrt{\widetilde g}
A \phi \left[{\wlBox}^2 
+ 2\widetilde R_{\mu\nu}\widetilde\nabla_\mu\widetilde\nabla_\nu 
 - {2 \over 3}\widetilde R \wlBox^2 
+ {1 \over 3}(\widetilde\nabla^\mu \widetilde R)\widetilde\nabla_\mu
\right]\phi \ .\nonumber
\eea 
Here the quantities in the  5 dimensional bulk spacetime are 
specified by the suffices $_{(5)}$ and those in the boundary 4 
dimensional spacetime are specified by $_{(4)}$. 
The factor $2$ in front of $S_1$ in (\ref{Stotal}) is coming from 
that we have two bulk regions which 
are connected with each other by the brane. 
In (\ref{GHi}), $n^\mu$ is 
the unit vector normal to the boundary. In (\ref{GHi}), (\ref{S1}) 
and (\ref{W}), one chooses 
the 4 dimensional boundary metric as 
\be
\label{tildeg}
\gfr_{\mu\nu}=\e^{2A}\tilde g_{\mu\nu},
\ee 
We should distinguish $A$ and $\tilde g_{\mu\nu}$ with 
$\tilde A(z)$ and $\hat g_{ij}$ in (\ref{AdS}). We will 
specify $\hat g_{ij}$ later in (\ref{metric1}). 
We also specify the 
quantities given by $\tilde g_{\mu\nu}$ by using $\tilde{\ }$. 
$G$ ($\tilde G$) and $F$ ($\tilde F$) are the Gauss-Bonnet
invariant and the square of the Weyl tensor, which are given as
\footnote{We use the following curvature conventions:
\begin{eqnarray*}
R&=&g^{\mu\nu}R_{\mu\nu} \nn
R_{\mu\nu}&=& R^\lambda_{\ \mu\lambda\nu} \nn
R^\lambda_{\ \mu\rho\nu}&=&
-\Gamma^\lambda_{\mu\rho,\nu}
+ \Gamma^\lambda_{\mu\nu,\rho}
- \Gamma^\eta_{\mu\rho}\Gamma^\lambda_{\nu\eta}
+ \Gamma^\eta_{\mu\nu}\Gamma^\lambda_{\rho\eta} \nn
\Gamma^\eta_{\mu\lambda}&=&{1 \over 2}g^{\eta\nu}\left(
g_{\mu\nu,\lambda} + g_{\lambda\nu,\mu} - g_{\mu\lambda,\nu} 
\right)\ .
\end{eqnarray*}}
\bea
\label{GF}
G&=&R^2 -4 R_{ij}R^{ij}
+ R_{ijkl}R^{ijkl}, \nn
F&=&{1 \over 3}R^2 -2 R_{ij}R^{ij}
+ R_{ijkl}R^{ijkl} \ ,
\eea

In the effective action (\ref{W}) induced by brane quantum matter,
 with $N$ scalar, $N_{1/2}$ spinor, $N_1$ vector fields, $N_2$ 
 ($=0$ or $1$) gravitons and $N_{\rm HD}$ higher 
derivative conformal scalars, $b$, $b'$ and $b''$ are 
\bea
\label{bs}
b&=&{N +6N_{1/2}+12N_1 + 611 N_2 - 8N_{\rm HD} 
\over 120(4\pi)^2}\nn 
b'&=&-{N+11N_{1/2}+62N_1 + 1411 N_2 -28 N_{\rm HD} 
\over 360(4\pi)^2}\ , 
\nn 
b''&=&0\ .
\eea
Usually, $b''$ may be changed by the finite 
renormalization of local counterterm in gravitational 
effective action. As it was the case in ref.\cite{NOO}, the term 
proportional 
to $\left\{b''+ {2 \over 3}(b + b')\right\}$ in (\ref{W}), and 
therefore $b''$, does not contribute to the equations of motion.
Note that CFT matter induced effective action may be considered as
brane dilatonic gravity.
 
For typical examples motivated by AdS/CFT correspondence\cite{AdS} 
one has:

\ 

\noindent
a) ${\cal N}=4$ $SU(N)$ SYM theory 
\be
\label{N4bb}
b=-b'={C \over 4}={N^2 -1 \over 4(4\pi )^2}\ ,
\ee 
b) ${\cal N}=2$ $Sp(N)$ theory 
\be
\label{N2bb}
b={12 N^2 + 18 N -2 \over 24(4\pi)^2}\ ,\quad 
b'=-{12 N^2 + 12 N -1 \over 24(4\pi)^2}\ .
\ee
One can write the corresponding expression for dilaton coupled spinor matter
\cite{peter} which also has non-trivial (slightly different in 
form) dilatonic contribution to CA than in case of holographic 
conformal 
anomaly\cite{LT} for ${\cal N}=4$ super Yang-Mills theory.

Let us consider the case where there are two branes at 
$z=\tilde z_0$ and $z=z_0$,  adding the action corresponding 
to the brane at $z=\tilde z_0$ to the action in (\ref{Stotal}):
\bea
\label{Stotal2}
S_{\rm two\ branes}&=&S+ \tilde\SGH + 2 \tilde S_1 + \tilde W, \\
\label{SEHib}
\tilde\SGH&=&{1 \over 8\pi G}\int d^4 x 
\sqrt{\gfr}\nabla_\mu n^\mu, \\
\label{S1b}
\tilde S_1&=& {1 \over 16\pi G l}\int d^4 x \sqrt{\gfr}\left(
{6 \over l} + {l \over 4}\Phi(\phi)\right), \\
\label{Wb}
\tilde W&=& \tilde b \int d^4x \sqrt{\widetilde g}\widetilde F A \\ 
&& \mbox{\hskip -1cm} + \tilde b' \int d^4x\sqrt{\widetilde g}
\left\{A \left[2{\wlBox}^2 
+\widetilde R_{\mu\nu}\widetilde\nabla_\mu\widetilde\nabla_\nu 
 - {4 \over 3}\widetilde R \wlBox^2 
+ {2 \over 3}(\widetilde\nabla^\mu \widetilde R)\widetilde\nabla_\mu
\right]A \right. \nn
&&  \mbox{\hskip -1cm}
\left. + \left(\widetilde G - {2 \over 3}\wlBox \widetilde R
\right)A \right\} \nn
&&  \mbox{\hskip -1cm} 
 -{1 \over 12}\left\{\tilde b''+ {2 \over 3}(\tilde b 
+ \tilde b')\right\}
\int d^4x \sqrt{\widetilde g} 
\left[ \widetilde R - 6\wlBox A 
 - 6 (\widetilde\nabla_\mu A)(\widetilde \nabla^\mu A)
\right]^2 \nn
&&  \mbox{\hskip -1cm} + \tilde C \int d^4x \sqrt{\widetilde g}
A \phi \left[{\wlBox}^2 
+ 2\widetilde R_{\mu\nu}\widetilde\nabla_\mu\widetilde\nabla_\nu 
 - {2 \over 3}\widetilde R \wlBox^2 
+ {1 \over 3}(\widetilde\nabla^\mu \widetilde R)\widetilde\nabla_\mu
\right]\phi \nonumber \ .
\eea 
We should note that the relative sign of $\tilde S_1$ is 
different from $S_1$. The parameters $\tilde b$, $\tilde b'$, 
$\tilde b''$ and $\tilde C$ correspond to the matter which may be 
different from the outer brane one on the inner brane as in (\ref{bs}). 
Hence, the situation with different CFTs on the branes may 
be considered.
Having the action at hands one can study its dynamics.

\section{Dilatonic quantum brane-worlds}

Let us start the consideration of field equations for two-branes model. 
First of all,
 one defines a new coordinate $z$ by
\be
\label{c2b}
z=\int dy\sqrt{f(y)},
\ee
and solves $y$ with respect to $z$. Then the warp
factor is $\e^{2\hat A(z,k)}=y(z)$. Here one assumes 
the metric of 5 dimensional spacetime as follows:
\be
\label{DP1}
ds^2=g_{(5)\mu\nu}dx^\mu dx^\nu =f(y)dy^2 
+ y\sum_{i,j=1}^4\hat g_{ij}(x^k)dx^i dx^j. 
\ee
Here $\hat g_{ij}$ is the metric of the 4 dimensional Einstein 
manifold as in (\ref{AdS}).  From the variation over $g_{(5)\mu\nu}$ 
in the Einstein-Hilbert 
action (\ref{SEHi}), we obtain the following equation in the 
bulk
\bea
\label{iit}
0&=&R_{(5)\mu\nu}-{1 \over 2}g_{(5)\mu\nu}R 
 - {1 \over 2}\left({l^2 \over 12} + \Phi(\phi)\right)
g_{(5)\mu\nu} \nn
&& - {1 \over 2} \left(\partial_\mu\phi\partial_\nu\phi 
 -{1 \over 2}g_{(5)\mu\nu}g_{(5)}^{\rho\sigma}\partial_\rho \phi
\partial_\sigma \phi \right)
\eea
and from that over dilaton $\phi$
\be
\label{iiit}
0=\partial_\mu\left(\sqrt{g_{(5)}}g_{(5)}^{\mu\nu}
\partial_\nu\phi\right) + \Phi'(\phi)\ .
\ee
%%%%%%%%%
Assuming that $g_{(5)\mu\nu}$ is given by (\ref{DP1}) and 
$\phi$ depends only on $y$: $\phi=\phi(y)$, we find 
the equations of motion (\ref{iit}) and (\ref{iiit}) take the 
following forms:
\bea
\label{DP2}
0&=&{2kf \over y}
 -{3 \over 2}{1 \over y^2} + {1 \over 2}
\left({l^2 \over 12} + \Phi(\phi)\right)f 
+ {1 \over 4}\left({d\phi \over dy}\right)^2 \\
\label{viitb}
0&=& {kf \over y} + {3 \over 4fy}{df \over dy}
+ {1 \over 2}\left({l^2 \over 12} + \Phi(\phi)\right)f
 - {1 \over 4}\left(d\phi \over dy\right)^2 \\
\label{DP3}
0&=&{d \over dy}\left({y^2 \over \sqrt{f}}{d\phi \over dy}\right)
+ \Phi'(\phi)y^2 \sqrt{f}\ .
\eea
Eq.(\ref{DP2}) corresponds to $(\mu,\nu)=(y,y)$ in (\ref{iit}) and 
Eq.(\ref{viitb}) to $(\mu,\nu)=(i,j)$. The case of $(\mu,\nu)=(y,i)$ 
or $(i,y)$ is identically satisfied. 

On the other hand, on the (outer) brane, we obtain the following 
equations:
\bea
\label{eq2b}
0&=&{48 l^4 \over 16\pi G}\left(\partial_z A - {1 \over l}
 - {l \over 24}\Phi(\phi)\right)\e^{4A}
+b'\left(4\partial_\sigma^4 A - 16 \partial_\sigma^2 A\right) \nn
&& - 4(b+b')\left(\partial_\sigma^4 A + 2 \partial_\sigma^2 A 
 - 6 (\partial_\sigma A)^2\partial_\sigma^2 A \right) \nn
&& + 2C\left(\partial_\sigma^4 \phi
 - 4 \partial_\sigma^2 \phi \right), \\
\label{eq2pb}
0&=&-{l^4 \over 8\pi G}\e^{4A}\partial_z\phi
 -{l^3 \over 32\pi G}\e^{4A}\Phi'(\phi) \nn
&& + C\left\{A\left(\partial_\sigma^4 \phi
 - 4 \partial_\sigma^2 \phi \right) 
+ \partial_\sigma^4 (A\phi)
 - 4 \partial_\sigma^2 (A\phi) \right\}\ .
\eea
For inner brane, one gets
\bea
\label{eq2bin}
0&=&-{48 l^4 \over 16\pi G}\left(\partial_z A - {1 \over l}
 - {l \over 24}\Phi(\phi)\right)\e^{4A}
+\tilde b'
\left(4\partial_\sigma^4 A - 16 \partial_\sigma^2 A\right) \nn
&& - 4(\tilde b+ \tilde b')\left(\partial_\sigma^4 A 
+ 2 \partial_\sigma^2 A 
 - 6 (\partial_\sigma A)^2\partial_\sigma^2 A \right) \nn
&& + 2\tilde C\left(\partial_\sigma^4 \phi
 - 4 \partial_\sigma^2 \phi \right), \\
\label{eq2pbin}
0&=&{l^4 \over 8\pi G}\e^{4A}\partial_z\phi
 +{l^3 \over 32\pi G}\e^{4A}\Phi'(\phi) \nn
&& + \tilde C\left\{A\left(\partial_\sigma^4 \phi
 - 4 \partial_\sigma^2 \phi \right) 
+ \partial_\sigma^4 (A\phi)
 - 4 \partial_\sigma^2 (A\phi) \right\}\ .
\eea
In (\ref{eq2b}) and (\ref{eq2pb}),  using 
the change of the coordinate: $dz=\sqrt{f}dy$ and 
 choosing $l^2\e^{2\hat A(z,k)}=y(z)$ one uses the form of 
the metric as 
\be
\label{metric1}
ds^2=dz^2 + \e^{2A(z,\sigma)}\tilde g_{\mu\nu}dx^\mu dx^\nu\ ,
\quad \tilde g_{\mu\nu}dx^\mu dx^\nu\equiv l^2\left(d \sigma^2 
+ d\Omega^2_3\right)\ .
\ee
Here $d\Omega^2_3$ corresponds to the metric of 3 dimensional 
unit sphere. Then for the unit sphere ($k=3$)
\be
\label{smetric}
A(z,\sigma)=\hat A(z,k=3) - \ln\cosh\sigma\ ,
\ee
for the flat Euclidean space ($k=0$)
\be
\label{emetric}
A(z,\sigma)=\hat A(z,k=0) + \sigma\ ,
\ee
and for the unit hyperboloid ($k=-3$)
\be
\label{hmetric}
A(z,\sigma)=\hat A(z,k=-3) - \ln\sinh\sigma\ .
\ee
We now identify $A$ and $\tilde g$ in (\ref{metric1}) with those in 
(\ref{tildeg}). Then we find $\tilde F=\tilde G=0$, 
$\tilde R={6 \over l^2}$ etc.

Using (\ref{DP2}) and (\ref{DP3}), one can delete $f$ from the 
equations and can obtain an equation that contains only the 
dilaton field $\phi$ (and, of course, bulk potential):
\bea
\label{DP4}
0&=&\left\{ {5k \over 2} - {k \over 4}y^2
\left({d\phi \over dy}\right)^2 + \left({3 \over 2}y 
 - {y^3 \over 6}\left({d\phi \over dy}\right)^2 \right)
\left({6 \over l^2} + {1 \over 2}\Phi(\phi)\right)\right\}
{d\phi \over dy} \nn
&& + {y^2 \over 2}\left({2k \over y} + {6 \over l^2} 
+ {1 \over 2}\Phi(\phi)\right){d^2\phi \over dy^2}
+ \left({3 \over 4} - {y^2 \over 8}
\left({d\phi \over dy}\right)^2 \right)\Phi'(\phi)\ .
\eea
%%%%%%%%%%%%%%%%%%%%%%%%%%%%%%%
Our choice for dilaton and bulk potential admitting the analytical 
solution is 
\bea
\label{assmp1}
\phi(y)&=&p_1\ln \left(p_2 y\right) \\
\label{assmp2}
\Phi(\phi)&=&-{12 \over l^2} + c_1 \exp\left(a\phi\right) 
+ c_2\exp\left(2a\phi\right)\ , 
\eea
where $a$, $p_1$, $p_2$, $c_1$, $c_2$ are some constants. 
When $p_1=\pm {1 \over \sqrt{6}}$, Eq.(\ref{DP4}) is always 
satisfied but from Eq.(\ref{DP3}), we find that $f(y)$ 
identically vanishes. Therefore we should assume $p_1\neq
\pm {1 \over \sqrt{6}}$. Then we find the following set of exact 
bulk solutions  
\bea
\label{case1}
&\mbox{case 1}\quad & c_1={6kp_2p_1^2 \over 3 - 2p_1^2}\ ,\quad 
c_2=0\ ,\quad a=-{1 \over p_1}\ ,\quad p_1\neq \pm \sqrt{6} \nn
&& f(y)={3- 2p_1^2 \over 4ky} \\ 
\label{case2}
&\mbox{case 2}\quad & c_1=-6kp_2\ ,\quad 
a=\pm{1 \over \sqrt{3}}\ ,\quad p_1=\mp\sqrt{3} \nn
&& f(y)={3 \over {2c_2 \over p_2^2} - 4ky} \\ 
\label{case3}
&\mbox{case 3}\quad & c_2=3kp_2\ ,\quad 
a=\pm{1 \over \sqrt{3}}\ ,\quad p_1=\mp{\sqrt{3} \over 2} \nn
&& f(y)={21\sqrt{p_2} \over 8\sqrt{y}\left(c_1y 
+ 7k\sqrt{p_2y}\right)} \ .
\eea
We can check that the above solutions satisfy (\ref{viitb}). 

In the coordinate system in (\ref{DP1}), 
Eq.(\ref{eq2pb}) for an outer brane has the following form:
\be
\label{eq2pc}
0=- {y_0^2 \over 8\pi G \sqrt{f(y_0)}}\partial_y\phi 
 - {y_0^2 \over 32\pi Gl}\Phi'(\phi_0) + 6C\phi_0\ ,
\ee
and (\ref{eq2bin}) for inner brane 
\be
\label{eq2pcin}
0= {\tilde y_0^2 \over 8\pi G \sqrt{f(\tilde y_0)}}\partial_y\phi 
  +{\tilde y_0^2 \over 32\pi Gl}\Phi'(\tilde\phi_0)
+ 6\tilde C\tilde\phi_0\ .
\ee
Here $\phi_0$ ($\tilde\phi_0$) is the value of the dilaton $\phi$ 
on the outer (inner) brane. 
We also find Eq.(\ref{eq2b}) for an outer brane 
has the following form:
\be
\label{eq2c1}
0= {3y_0^2 \over 16\pi G}\left({1 \over 2y_0\sqrt{f(y_0)}}
 - {1 \over l} - {l \over 24}\Phi(\phi_0)\right) + 8b'
\ee
for $k\neq 0$ case and  
\be
\label{eq2c2}
0= {3y_0^2 \over 16\pi G}\left({1 \over 2y_0\sqrt{f(y_0)}}
 - {1 \over l} - {l \over 24}\Phi(\phi_0)\right) 
\ee
for $k=0$ case. For the inner brane (\ref{eq2bin}) for $k\neq 0$ 
has the form of 
\be
\label{eq2c1in}
0= -{3\tilde y_0^2 \over 16\pi G}\left({1 \over 2\tilde y_0
\sqrt{f(\tilde y_0)}}
 + {1 \over l} + {l \over 24}\Phi(\tilde \phi_0)\right) 
 + 8\tilde b'\ .
\ee
The equation for $k=0$ case is identical with that of the outer 
brane in (\ref{eq2c2}) if we replace $\tilde b'$ with $b'$. 

\ 

\noindent
{\bf Case 1 solution}

First we consider case 1 in (\ref{case1}). Since $f(y)$ should 
be positive (we should also note $y>0$), one gets 
\be
\label{qq}
q^2\equiv {4k \over 3 - 2p_1^2}>0\ ,\quad (q>0)\ .
\ee
In (\ref{qq}), we can also consider the limit of 
$k\rightarrow 0$ by keeping $q$ finite, i.e., 
$p_1^2\rightarrow {3 \over 2}$. 

When $k\neq 0$, Eqs.(\ref{eq2c1}) and 
(\ref{eq2pc}) have the following form:
\bea
\label{eq2c11}
-8b'&=&F_1(y_0)\equiv {3 \over 16\pi G}\left(
{q \over 2}y_0^{3 \over 2} - {1 \over 2l}y_0^2
 - {q^2p_1^2ly_0 \over 16}\right) \nn
&=& - {3 \over 16\pi G}{y_0 \over 2l}
\left( y_0^{1 \over 2} - {1 + \sqrt{1 - {p_1^2 l^2 \over 2}}
\over 2}q\right)
\left( y_0^{1 \over 2} - {1 - \sqrt{1 - {p_1^2 l^2 \over 2}}
\over 2}q\right) \\
\label{eq2pc1}
0&=& - {p_1q \over 8\pi G}y_0^{3 \over 2} 
+ {3lp_1 q^2 \over 64\pi G}y_0 + 6C\phi_0\ .
\eea
and Eqs.(\ref{eq2bin}) and 
(\ref{eq2pbin}) are
\bea
\label{eq2c11in}
8\tilde b'&=&F_1(y_0)\equiv {3 \over 16\pi G}\left(
{q \over 2}y_0^{3 \over 2} - {1 \over 2l}y_0^2
 - {q^2p_1^2ly_0 \over 16}\right) \\
\label{eq2pc1in}
0&=& {p_1 q \over 8\pi G}y_0^{3 \over 2} 
 - {3lp_1 q^2 \over 64\pi G}y_0 + 6\tilde C\tilde\phi_0\ .
\eea
Since $p_2$ is absorbed into the definition of $q$ in 
(\ref{eq2c11}) and (\ref{eq2c11in}), Eqs.(\ref{eq2pc1}) 
and (\ref{eq2pc1in}) can be regarded as the equation 
which determines $p_2$ 
or ${\phi_0 \over p_1}\ln\left(p_2 y_0\right)$
and ${\tilde \phi_0 \over p_1}=\ln\left(p_2 \tilde y_0\right)$. 
We now investigate the properties of $F_1(y_0)$ as a function 
of $y_0$. The asymptotic behaviors are given by
\bea
\label{F1}
&F_1(y_0)\rightarrow - {3 \over 16\pi G}\cdot 
{p_1^2  q^2 l \over 16}y_0 < 0\quad &\mbox{when}\ 
y_0 \rightarrow +0 \\
\label{F2}
&F_1(y_0)\rightarrow - {3 \over 16\pi G}\cdot 
{1 \over 2l}y_0^2 < 0\quad &\mbox{when}\ 
y_0 \rightarrow +\infty \ .
\eea
Since 
\be
\label{F3}
F'_1(y_0)= {3 \over 16\pi G}\left( {3q \over 4}y^{1 \over 2}
 - {1 \over l}y_0 - {p_1^2 q^2 l \over 16}\right)\ ,
\ee
$F_1(y_0)$ has extrema when
\be
\label{F4}
0=y_0 - {3ql \over 4}y_0^{1 \over 2} 
+ {p_1^2 q^2 l^2 \over 16}\ ,
\ee
whose solutions are given by
\be
\label{F5}
y_0^{1 \over 2}=y_\pm^{1 \over 2}
\equiv {3ql \over 8}\left(1\pm \sqrt{1 
- {4p_1^2 \over 9}}\right)\ .
\ee
Therefore if 
\be
\label{F6}
|p_1|>{3 \over 2}\ ,
\ee
Eq.(\ref{F4}) does not have any solution and $F_1(y_0)$ 
is monotonically decreasing function of $y_0$. Then 
Eqs.(\ref{F1}) and (\ref{F2}) tell that there is no solution 
of the brane equation (\ref{eq2c11}) for negative $b'$ in 
case of (\ref{F6}). 
On the other hand, when 
\be
\label{F6b}
|p_1|<{3 \over 2}\ ,
\ee
 substituting (\ref{F5}) into the expression for $F_1(y_0)$ in 
(\ref{eq2c11}), one gets
\be
\label{F7}
F_1(y_\pm)={3 \over 16\pi G}{3^4 q^4 l^3 \over 2\cdot 8^4}
\left(\sqrt{1-{4p_1^2 \over 9}} \pm 1\right)
\left(\sqrt{1-{4p_1^2 \over 9}} \mp {1 \over 3}\right)\ .
\ee
Then we find $y_0=y_+$ corresponds to the maximum of $F_1(y_0)$. 
The maximum is positive $F_1(y_+)>0$ if $\sqrt{1-{4p_1^2 \over 9}} 
 - {1 \over 3}>0$, that is,
\be
\label{F8}
p_1^2<2\ ,
\ee
which is, of course, consistent with (\ref{F6b}). 
In case of (\ref{F8}), if 
\be
\label{F9}
F_1(y_+)\geq -8b'\ , 
\ee
Eq.(\ref{eq2c11}) has a solution, that is, there can be a 
brane. We can also consider an inner brane which lies at 
$y=y_1<y_0$. For the inner brane, the relative sign of 
$\tilde b'$ and $b'$ 
is changed in the equation corresponding to (\ref{eq2c11}). 
Then if 
\be
\label{F10}
F_1(y_-)\leq 8\tilde b'\ ,
\ee
there can be an inner brane. Then if both of (\ref{F9}) 
and (\ref{F10}) hold, we can have two brane dilatonic solution. 
In case of two brane solution, there might be, in general, 
a problem in the consistency between (\ref{eq2pc1}) and 
(\ref{eq2pc1in}). If we impose both of (\ref{eq2pc1}) and 
(\ref{eq2pc1in}), they can be regarded as the equations 
which determine $p_1$ and $p_2$ (we should note that $p_2$ 
is implicitly contained in $\phi_0$ and $\tilde\phi_0$). 
In the classical limit, where $C=0$, there disappear the 
terms containing $p_2$ (or $\phi_0$ and $\tilde\phi_0$). 
Then it seems to be non-trivial if there exists any solution 
which satisfies both of (\ref{F9}) and (\ref{F10}). 

We now consider the classical limit in $k\neq 0$ case, where 
$b'=C=\tilde b'=\tilde C=0$ and (\ref{eq2c11}) and (\ref{eq2c11in}) become 
identical. Then the solutions of Eqs.(\ref{eq2c11}) and 
(\ref{eq2c11in})  are given by 
\be
\label{cc1a}
y_0^{1 \over 2} = \left(1 \pm \sqrt{1 - {p_1^2 \over 2}}
\right){ql \over 2}\ .
\ee
Since both of the solutions are positive, we can regard 
smaller one ($-$ sign in (\ref{cc1a})) as expressing inner 
brane and larger one ($+$ sign) as an outer brane. 
On the other hand, Eqs.(\ref{eq2pc1}) and (\ref{eq2pc1in}) 
have the following form:
\be
\label{cc1b}
0={p_1 q^2 l y_0 \over 2}\left(-{1 \over 4}\mp \sqrt{1
 - {p_1^2 \over 2}}\right)\ .
\ee
In (\ref{cc1b}), 
the upper sign ($-$) corresponds to the outer brane and the lower 
one ($+$) to the inner brane. We should note that there is no 
solution, except for an outer brane $p_1^2={15 \over 8}$. 
This would tell 
that we need the quantum correction from brane matter in order to obtain the 
two brane dilatonic inflationary Universe where observable world 
may be associated with one of inflationary branes. 

%%%%%%%%%%%%%%%%%%%%%

When $k=0$ in case 1, as discussed before, if $q$ is 
finite, we find 
\be
\label{c1k0i}
p_1^2\rightarrow {3 \over 2}\ .
\ee
Then Eq.(\ref{eq2c2}) can be rewritten in the following form:
\be
\label{c1k0ii}
0=y_0 - ql y_0^{1 \over 2} + {3q^2l^2 \over 16}\ ,
\ee
which has two solutions:
\be
\label{c1k0iii}
y_0^{1 \over 2}={3ql \over 4}\ ,\quad {ql \over 4}\ .
\ee
These two solutions might be regarded as two brane solutions. 
On the other hand, the form of Eq.(\ref{eq2pc}) for $k=0$ case 
is identical with that of $k\neq 0$ in (\ref{eq2pc1}), which can 
be solved with respect to $\phi_0$ or $p_2$ in one brane solution. 
However, in case $k=0$, the value of $p_1$ is fixed by 
(\ref{c1k0i}). In the classical limit of $k=0$  case, 
Eq.(\ref{eq2c2}) can be rewritten in the form of (\ref{c1k0ii}) 
and there appear solutions in (\ref{c1k0iii}). Eq.(\ref{eq2pc}) 
is, however, not satisfied. Eq.(\ref{eq2pc}) has a form of 
(\ref{cc1b}) but Eq.(\ref{c1k0i}) does not satisfy (\ref{cc1b}).
This demonstrates the role of quantum effects in realization
of dilatonic inflationary two brane-world Universe.

\ 

\noindent
{\bf Case 2 solution}

We now consider the case 2 in (\ref{case2}). Defining 
$\tilde c_2$ as
\be
\label{tildec2}
\tilde c_2\equiv {c_2 \over p_2^2}\ ,
\ee
Eqs.(\ref{eq2c1}) and (\ref{eq2pc}) for the outer brane 
have the following form ( when $k\neq 0$):
\bea
\label{eq2c21}
-8b'&=&F_2(y_0)\equiv {3 \over 16\pi G}\left(
{y_0 \over 2}\sqrt{2\tilde c_2 - 4ky_0 \over 3}
 - {y_0^2 \over 2l} + {kly_0 \over 4}
 - {l \tilde c_2 \over 24}\right) \\
\label{eq2pc2}
0&=& - {y_0 \over 8\sqrt{3}\pi G}\sqrt{2\tilde c_2 - 4ky_0 \over 3}
- {l y_0^2 \over 32\pi G}\left( - {6k \over \sqrt{3} y_0} 
+ {2 \tilde c_2 \over \sqrt{3}y_0^2}\right) + 6C\phi_0\ ,
\eea
and Eqs.(\ref{eq2c1in}) and (\ref{eq2pcin}) for the inner 
brane, when $k\neq 0$:
\bea
\label{eq2c21in}
8\tilde b'&=&F_2(\tilde y_0) \\
\label{eq2pc2in}
0&=& {\tilde y_0 \over 8\sqrt{3}\pi G}
\sqrt{2\tilde c_2 - 4k\tilde y_0 \over 3}
+ {l \tilde y_0^2 \over 32\pi G}
\left( - {6k \over \sqrt{3} \tilde y_0} 
+ {2 \tilde c_2 \over \sqrt{3}\tilde y_0^2}\right) 
+ 6\tilde C\tilde\phi_0\ ,
\eea
%%%%%%%%%%%%%%%%%%%%%%%%%%%%%%
Since $p_2$ is absorbed into the definition of $\tilde c_2$ in 
(\ref{eq2c21}) and (\ref{eq2c21in}), Eqs.(\ref{eq2pc2}) 
and (\ref{eq2pc2in}) can be regarded again 
as the equation which determines $p_2$ or 
$\phi_0=p_1\ln\left(p_2 y_0\right)$ ($\tilde\phi_0
=p_1\ln\left(p_2 \tilde y_0\right)$. 
Then 
\be
\label{F2a}
F_2'(y_0)= {3 \over 16\pi G}\left(
{1 \over 2}\sqrt{2\tilde c_2 - 4ky_0 \over 3}
 -{{ky_0 \over 3} \over \sqrt{2\tilde c_2 - 4ky_0 \over 3}}
 - {y_0 \over l} + {kl \over 4}\right)\ .
\ee
Then if $F_2'(y_0)=0$, one gets 
\be
\label{F2b}
0=f(y_0)\equiv {4k \over l^2}y_0^3 + \left(k^2 
 - {2\tilde c_2 \over l^2}\right)y_0^2 
+ \left({k^3l^2 \over 4} - \tilde c_2 k\right)y_0 
+\left( -{k^2 l^2 \over 8} + {1 \over 3}\right)\tilde c_2
\ee
and
\be
\label{F2c}
f'(y_0)={12k \over l^2}y_0^2 + 2\left(k^2 
 - {2\tilde c_2 \over l^2}\right)y_0 
+ \left({k^3l^2 \over 4} - \tilde c_2 k\right)\ .
\ee
Then if we further put $f'(y_0)=0$, the determinant $D$ of the 
equation is given by
\bea
\label{F2d}
{D \over 4}&=&\left(k^2 - {2\tilde c_2 \over l^2}\right)^2
 - {12k \over l^2}\left({k^3l^2 \over 4}
 - \tilde c_2 k\right) \nn
&=&{4 \over l^2}\left\{\tilde c_2 + k^2l^2 \left(1 
+ \sqrt{3 \over 2} \right)\right\}\left\{\tilde c_2 + k^2l^2 \left(1 
 - \sqrt{3 \over 2} \right)\right\} \ .
\eea
If $D<0$, the equation $f'(y_0)=0$ does not have any solution. 
Then there can be only one solution $f(y_0)=0$, then in this case, 
$F_2(y_0)$ can have only one extremum. 
The explicit solutions of (\ref{F2b}) are given by
\bea
\label{F2e}
y_0&=&-{l^2k(1 - 2\hat c)  \over 12k}
+ \left(-\beta + \sqrt{\beta^2 + \alpha^3}
\right)^{1 \over 3}\omega
+ \left(-\beta - \sqrt{\beta^2 + \alpha^3}
\right)^{1 \over 3}\omega^2 \nn
\omega&=&1,\ \e^{2\pi i \over 3},\ \e^{4\pi i \over 3} \nn
\alpha&\equiv&{2 - 8\hat c - 4\hat c^2 \over 3} \nn
\beta&\equiv&{-7 - 12\hat c + 96\hat c^2 - 16\hat c^3 
\over 54} \nn
\hat c&=&{\tilde c_2 \over k^2 l^2} \ .
\eea
Then if 
\be
\label{F2f}
\beta^2 + \alpha^3<0\ ,
\ee
Eq.(\ref{F2b}) has three different real solutions, and  
$F_2(y_0)$ can have three extrema (at maximum).

Let us consider the solution of Eq.(\ref{eq2c21}) or the behavior 
of $F_2(y_0)$ in more detail. 

In case of $k>0$, since $F_2(y_0)$ contains 
$\sqrt{2\tilde c_2 - 4ky_0 \over 3}$, the value of 
$y_0$ is restricted to be $0\leq y_0\leq {\tilde c_2 \over 2k}$ 
and $\tilde c_2$ should be positive: $\tilde c_2>0$. Since 
\bea
\label{F2g}
F_2(0)&=& - {3 \over 16\pi G}{l\tilde c_2 \over 24}<0 \nn
F_2\left({\tilde c_2 \over 2k}\right)
&=& - {3 \over 16\pi G}{1 \over 8k^2 l}
\left(\tilde c_2 - {2k^2 l \over 3}\right)\tilde c_2 \ ,
\eea
Eq.(\ref{eq2c21}) has an outer brane solution, at least, if 
$F_2\left({\tilde c_2 \over 2k}\right)\geq -8b'$. As usually 
$-8b'>0$, $F_2\left({\tilde c_2 \over 2k}\right)$ should be 
positive, which requires $\tilde c_2 < {2k^2 l \over 3}$. 
More generally, since $F'_2\left(0\right)>0$ and 
$F'_2\left(y_0\rightarrow 
{\tilde c_2 \over 2k}\right)\rightarrow -\infty$, 
$F_2\left(y_0\right)$ can have at least one maximum. If the 
maximum is greater than $-8b'$ ($=0$ in the classical case), 
there can be an outer brane solution(s). And if $F_2(0)<8\tilde b'$ 
($=0$ in the classical case), there can be an inner brane 
solution(s). We should note that such an inner and/or outer brane(s) 
soultion(s) can exist in general even if $b'=\tilde b'=0$.    
Hence, the possibility of creation of inflationary two brane-world 
Universe occurs not only on quantum but also on classical level 
(depending on the choice of the parameters).

We now consider the case of $k<0$ for (\ref{eq2c21}) 
in the case 2. If $\tilde c_2>0$, $y_0$ can take a value 
from $0$ to positive infinity: $0\leq y_0 < \infty$. 
Since 
\bea
\label{F2h}
&& F_2(0)= - {3 \over 16\pi G}{l\tilde c_2 \over 24}<0\ , 
\quad F'_2(0)= { 3 \over 16\pi G}\left(
\sqrt{2\tilde c_2 \over 6} + {kl \over 4}\right) \nn
&& F_2\left(y_0\rightarrow + \infty \right)\rightarrow 
- {3 \over 16\pi G}{y_0^2 \over 2l}<0 \ ,
\eea
if $\tilde c_2> {3k^2l^2 \over 8}$, $F_2(y_0)$ has at least one 
maximum. If the value of the maximum is larger than $-8b'$ ($=0$ 
in the classical case), there 
is always an outer brane solution. Even if $\tilde c_2< 
{3k^2l^2 \over 8}$, from (\ref{F2d}), there can be a maximum 
when $\tilde c_2 > k^2l^2 \left( \sqrt{3 \over 2} -1\right)$. 
Here we should note ${3 \over 8}> \sqrt{3 \over 2} -1$. 
When $\tilde c_2 < k^2l^2 \left( \sqrt{3 \over 2} -1\right)$, 
$F_2(y_0)$ becomes a monotonically decreasing function of $y_0$. 
Since $F_2(0)$ is negative, there cannot be any outer brane 
solution. 

In case that $k<0$ and $\tilde c_2<0$, we find $y_0> 
{\tilde c_2 \over 2k}$ in order that $f(y)$ is positive.
The surface of $y_0= {\tilde c_2 \over 2k}$ can be regarded as 
a horizon. Since 
\bea
\label{F2h2}
F_2\left({\tilde c_2 \over 2k}\right)
&=& - {3 \over 16\pi G}{1 \over 8k^2 l}
\left(\tilde c_2 - {2k^2 l \over 3}\right)\tilde c_2<0 \nn
&& F_2\left(y_0\rightarrow + \infty \right)\rightarrow 
- {3 \over 16\pi G}{y_0^2 \over 2l}<0 \ ,
\eea
and $F_2\left(y_0\rightarrow 
{\tilde c_2 \over 2k}\right)\rightarrow +\infty$, there is at 
least one maximum. If the maximum is larger then $-8b'$ (when 
$b'<0$), there are outer brane solutions. And if 
$F_2\left({\tilde c_2 \over 2k}\right)<8\tilde b'$, there can be an 
inner brane solution.

Finally, when $k=0$ in case 2 (\ref{case2}), the brane 
equation (\ref{eq2c2}) has the following form:
\be
\label{F20a}
0=y_0^2 - ly_0\sqrt{2\tilde c_2 \over 3} 
+ {2l^2\tilde c_2 \over 3}\ .
\ee
Then $\tilde c_2$ should be positive. The solution of 
(\ref{F20a}) is given by 
\be
\label{F20b}
y_0={l\sqrt{\tilde c_2} \over 2}\left(\sqrt{2 \over 3} \pm 
{1 \over \sqrt{3}}\right)\ .
\ee
Since both of the two solutions are positive, there can be 
a solution with both of inner and outer branes. 

\ 

\noindent
{\bf Case 3 solution}

We now briefly consider case 3 (\ref{case3}). 
First one should note that $p_2>0$ since the solution
(\ref{case3}) contains $\sqrt{p_2}$. Then if we define 
$\tilde c_1$ by
\be
\label{F3a}
\tilde c_1\equiv {c_1 \over \sqrt{p_2}}\ ,
\ee
when $k\neq 0$, Eqs.(\ref{eq2c1}) and 
(\ref{eq2pc}) have the following form:
\bea
\label{eq2c13}
-8b'&=&F_3(y_0) \nn
&\equiv& {3 \over 16\pi G}\left\{
{y_0 \over 2y_0}\sqrt{8\sqrt{y_0}\left(\tilde c_1 y_0 
+ 7k\sqrt{y_0}\right) \over 21} \right.\nn
&& \left. - {y_0^2 \over 2l} - {l \over 24}
\sqrt{y_0}\left(\tilde c_1 y_0 + 3k\sqrt{y_0}\right)
\right\} \\
\label{eq2pc3}
0&=& {y_0 \over 16\pi G}\sqrt{2\sqrt{y_0}\left(\tilde c_1 y_0 
+ 7k\sqrt{y_0}\right) \over 7} \nn
&& -{l\sqrt{3y_0} \over 96\pi G}\left(3k p_2\tilde c_1 y_0 
+ {2c_2\sqrt{y_0} \over p_2}\right) 
 - 3\sqrt{3}C\phi_0\ .
\eea
Since $p_2$ is absorbed into the definition of $\tilde c_1$ in 
(\ref{eq2c13}), Eq.(\ref{eq2pc3}) can be regarded as the equation 
which determines $p_2$ or $\phi_0=p_1\ln\left(p_2 y_0\right)$. 

When $\tilde c_1$ is negative, we find $k>0$ and 
$0<y_0<{49 k^2 \over \tilde c_1^2}$ in order that $F_3(y_0)$ 
is real. Since 
\bea
\label{F3b}
F_3\left(y_0\rightarrow 0\right)
&=& -{3 \over 16\pi G} {lky_0 \over 8} <0 \nn
F_3\left({49 k^2 \over \tilde c_1^2}\right)&=& {3 \over 16\pi G}
{(7k)^3 \over \tilde c_1^2}\left({l \over 42} - {7k \over 2l
\tilde c_1^2}\right)\ .
\eea
Then if $F_3\left({49 k^2 \over \tilde c_1^2}\right)>-8b'$ ($=0$ 
in the classical case), there can be an outer brane solution. 

When $\tilde c_1$ is positive, $y_0$ can take a value from $0$ 
to $+\infty$ if $k$ is positive. Since 
\be
\label{F3c} 
F_3\left(y_0\rightarrow 0\right)
= -{3 \over 16\pi G} {lky_0 \over 8} <0 \ ,\quad 
F_3\left(y_0\rightarrow +\infty \right)
= -{3 \over 16\pi G} {y_0^2 \over 2l} <0 \ ,
\ee
it is not so clear if there can be any outer brane solution. 

When $\tilde c_1>0$ and $k<0$, we find 
$y_0>{49 k^2 \over \tilde c_1^2}$ in order that $F_3(y_0)$ 
is real. Since 
\bea
\label{F3d}
F_3\left({49 k^2 \over \tilde c_1^2}\right)&=& {3 \over 16\pi G}
{(7k)^3 \over \tilde c_1^2}\left({l \over 42} - {7k \over 2l
\tilde c_1^2}\right) > 0 \nn
F_3\left(y_0\rightarrow +\infty \right)
&=& -{3 \over 16\pi G} {y_0^2 \over 2l} <0 \ ,
\eea
there always exists an outer brane solution if 
$F_3\left({49 k^2 \over \tilde c_1^2}\right)>-8b'$.

\ 

We now summarize the obtained results. Generally the obtained 
bulk solutions have the form (the transformation of metric is discussed below)
\bea
\label{assmp1b}
\phi(y)&=&p_1\ln \left(p_2 y\right) \nn
\label{assmp2b}
\Phi(\phi)&=&-{12 \over l^2} + c_1 \exp\left(a\phi\right) 
+ c_2\exp\left(2a\phi\right)\ .
\eea
\begin{enumerate}
\item Case 1
\begin{enumerate}
\item bulk solution
\bea
\label{case1b}
&& c_1={6kp_2p_1^2 \over 3 - 2p_1^2}\ ,\quad 
c_2=0\ ,\quad a=-{1 \over p_1}\ ,\quad p_1\neq \pm \sqrt{6} \nn
&& f(y)={3- 2p_1^2 \over 4ky} \ .
\eea
\item When $k\neq 0$ and $p_1^2<2$, there is an outer brane 
solution if
\be
\label{F9b}
F_1(y_+)\geq -8b'\ , 
\ee
and there is an inner brane solution if
\be
\label{F10b}
F_1(y_-)\leq 8\tilde b'\ .
\ee
Here $F_1$ is defined by
\be
\label{F1b}
F_1(y_0)\equiv {3 \over 16\pi G}\left(
{q \over 2}y_0^{3 \over 2} - {1 \over 2l}y_0^2
 - {q^2p_1^2ly_0 \over 16}\right) 
\ee
and $y_\pm$ is given by
\be
\label{ypm}
y_\pm^{1 \over 2}
\equiv {3ql \over 8}\left(1\pm \sqrt{1 
- {4p_1^2 \over 9}}\right)\ .
\ee
\item Solution for $k=0$
\be
\label{c1k0ib}
p_1^2\rightarrow {3 \over 2}\ ,\quad 
y_0^{1 \over 2}={3ql \over 4}\ ,\quad {ql \over 4}\ .
\ee
\end{enumerate}
\item Case 2
\begin{enumerate}
\item bulk solution
\bea
\label{case2b}
&& c_1=-6kp_2\ ,\quad 
a=\pm{1 \over \sqrt{3}}\ ,\quad p_1=\mp\sqrt{3} \nn
&& f(y)={3 \over {2c_2 \over p_2^2} - 4ky} \ . 
\eea
\item In case of $k>0$, $\tilde c_2\equiv {c_2 \over p_2^2}$ 
should be positive and there is an outer brane solution, at 
least if  $F_2\left({\tilde c_2 \over 2k}\right)\geq -8b'$, where 
\be
\label{F2bbb}
F_2(y_0)\equiv {3 \over 16\pi G}\left(
{y_0 \over 2}\sqrt{2\tilde c_2 - 4ky_0 \over 3}
 - {y_0^2 \over 2l} + {kly_0 \over 4}
 - {l \tilde c_2 \over 24}\right) \ .
\ee
\item In case of $k<0$, $F_2(y)$ has at least one mimimum 
if $\tilde c_2 < k^2l^2 \left( \sqrt{3 \over 2} -1\right)$ 
or $\tilde c_2>0$. If the value of $F_2(y)$ at the maximum is 
larger than $-8b'$, there is an outer brane solution. If 
$\tilde c_2>0$ and $F_2(0)<8\tilde b'$ or 
$\tilde c_2<0$ and $F_2\left({\tilde c_2 \over 2k}\right)
<8\tilde b'$, 
there can be an inner brane solution. 
\item In case of $k=0$, if $\tilde c_2>0$, 
the solution is given by
\be
\label{F20bb}
y_0={l\sqrt{\tilde c_2} \over 2}\left(\sqrt{2 \over 3} \pm 
{1 \over \sqrt{3}}\right)\ .
\ee
\end{enumerate}
\item Case 3
\begin{enumerate}
\item bulk solution
\bea
\label{case3b}
&& c_2=3kp_2\ ,\quad 
a=\pm{1 \over \sqrt{3}}\ ,\quad p_1=\mp{\sqrt{3} \over 2} \nn
&& f(y)={21\sqrt{p_2} \over 8\sqrt{y}\left(c_1y 
+ 7k\sqrt{p_2y}\right)} \ .
\eea
\item When $\tilde c_1\equiv {c_1 \over \sqrt{p_2}}<0$, $k>0$ and 
there can be outer brane solution if 
$F_3\left({49 k^2 \over \tilde c_1^2}\right)>-8b'$, where
\bea
\label{F3bbb}
F_3(y_0) &\equiv& {3 \over 16\pi G}\left\{
{y_0 \over 2y_0}\sqrt{8\sqrt{y_0}\left(\tilde c_1 y_0 
+ 7k\sqrt{y_0}\right) \over 21} \right.\nn
&& \left. - {y_0^2 \over 2l} - {l \over 24}
\sqrt{y_0}\left(\tilde c_1 y_0 + 3k\sqrt{y_0}\right)
\right\}\ .
\eea
\item When $\tilde c_1>0$ and $k<0$, 
there always exists outer brane solution if 
$F_3\left({49 k^2 \over \tilde c_1^2}\right)>-8b'$.
\end{enumerate}
\end{enumerate}

%%%%%
 From the above results in case 1$\sim$3, we find there very often 
appear two (inner and outer) branes solution as in the first model 
by Randall and Sundrum \cite{RS1}. Moreover, the branes may be curved 
as de Sitter or hyperbolic space which gives the  way for ever 
expanding inflationary Universe. Such 
solutions often can exist even if there is no any quantum effect, i.e., 
$b'=0$. 

Let us make few remarks on the form of metric.
If one considers the metric in the form (\ref{AdS}), 
 the warp factor $\e^{2\tilde A(z)}$ does not behave as 
an exponential function of $z$ but as a power of $z$. 
For example, in case 1 (\ref{case1}), 
\bea
\label{c1metric} 
ds^2&=&{dy^2 \over q^2y} + y\sum_{i,j=1}^4\hat g_{ij}dx^i dx^j \nn
&=&dz^2 + {z^2 \over 4q}\sum_{i,j=1}^4\hat g_{ij}dx^i dx^j\ 
\eea
where $z=2q\sqrt{y}$. In case 2 (\ref{case2})
\bea
\label{c2metric}
ds^2&=&{3dy^2 \over 2\tilde c - 4ky} 
+ y\sum_{i,j=1}^4\hat g_{ij}dx^i dx^j \nn
&=&dz^2 + \left({\tilde c_2 \over 2}
 - {2k^2z^2 \over 3k}\right)\sum_{i,j=1}^4\hat g_{ij}dx^i dx^j\ ,
\eea
where $z={1 \over k}\sqrt{{3\tilde c_2 \over 2} - 3ky}$. One should 
note that the exponential behavior of the warp 
factor $\e^{A(z)}\sim \e^{\gamma z}$ requires $f(y)\sim 
{1 \over y^2}$, which tells that the spacetime is nearly AdS:
\bea
\label{expm}
ds^2&\sim &{dy^2 \over \gamma^2y^2} 
+ y\sum_{i,j=1}^4\hat g_{ij}dx^i dx^j \nn
&=& dz^2 + \e^{\gamma z}\sum_{i,j=1}^4\hat g_{ij}dx^i dx^j\ ,
\eea
where $y=\e^{\gamma z}$. 
This would require that we need a region (of complete spacetime) 
where, the potential and the dilaton become almost constant. 
It results in difficulties when one tries to explain the hierarchy 
using this model. 

Hence, we presented number of dilatonic (inflationary, flat or 
hyperbolic) two brane-world Universes which are created by 
quantum effects of brane matter. Sometimes, such Universes may be 
realized due to specific choice of dilatonic potential even 
on classical level.

\section{Multi-brane generalization}

In some papers (for example in \cite{HSTT}), the solution with many branes 
was proposed. In such model, there are two AdS spaces with the different 
radii or different values of the cosmological constants. They are 
glued by a brane, whose 
tension is given by the difference of the inverse
of the radii. In the solution, the value of ${dA \over dz}$ in 
the metric of the form (\ref{AdS}) jumps at the brane, which 
tells the value of $f(y)$ in the metric choice in (\ref{DP1}) 
jumps on the brane since $\sqrt{f(y)}={dz \over dy}
={1 \over 2y {dA \over dz}}$. 
Imagine one includes the quantum effects on the brane. Then one can, 
in general, glue  two AdS-like spaces with same values of 
the cosmological constant. Let us assume that there is a brane at 
$y=\hat y_0$ and there are two AdS-like spaces in $y>\hat y_0$ 
and $y<\hat y_0$ glued by the brane. One now denotes the quantity 
in the AdS-like space in $y>\hat y_0$ ($y<\hat y_0$) by the 
suffix $+$ ($-$). If we consider the case where the value of 
$l$ is identical in two AdS-like space and the value of the 
dilaton is continuous at the brane, we need not the counter term 
corresponding to $S_1$ in (\ref{S1}) or $\tilde S_1$ 
in (\ref{S1b}) since the action corresponding to $S_1$ cancells 
with $\tilde S_1$ (note that the relative sign between 
$S_1$ and $\tilde S_1$ is opposite). Then instead of 
(\ref{eq2pc}) and (\ref{eq2c1}) or 
(\ref{eq2pcin}) and (\ref{eq2c1in}), one obtains 
\bea
\label{eq2pcM}
0&=&- {\hat y_0^2 \over 8\pi G} \left(
{\partial_y\phi_+(\hat y_0) \over \sqrt{f_+(\hat y_0)}}
 - {\partial_y\phi_-(\hat y_0) \over \sqrt{f_-(\hat y_0)}}\right) 
+ 6\hat C\hat\phi_0(\hat y_0) \\
\label{eq2c1M}
0&=& {3\hat y_0 \over 16\pi G}\left({1 \over 2\sqrt{f_+(\hat y_0)}}
-{1 \over 2\sqrt{f_-(\hat y_0)}}\right) + 8\hat b'\ 
\eea
for $k\neq 0$ cases. Here we denote the quantities on the brane 
by $\hat{}$. For $k\neq 0$, we cannot put a brane without 
making the cosmological constants in the AdS-like spaces 
$y>\hat y_0$ and $y<\hat y_0$ different as in the case of  
\cite{HSTT} where no quantum corrections case has been considered. 

As an example, we only limit to case 1 in (\ref{case1})
\be
\label{case1M}
f_\pm = {1 \over q_\pm^2 y}\ .
\ee
Then  using (\ref{eq2c1M}), one finds 
\be
\label{eq2c1M2}
0= {3\sqrt{\hat y_0} \over 16\pi G}\left({1 \over 2q_+}
-{1 \over 2q_-}\right) + 8\hat b'\ .
\ee
On the other hand, from (\ref{eq2pcM}), we obtain  
\be
\label{eq2pcM2}
0=- {\hat y_0^{3 \over 2} \over 8\pi G }\left(
q_+p_{1+} - q_-p_{1-} \right) + 6\hat C \phi(\hat y_0) \ .
\ee
The condition of the continuity of the dilaton field at 
the brane gives, from (\ref{assmp1}), 
\be
\label{M3}
\phi(\hat y_0) = p_{1+}\ln (p_{2+}\hat y_0)
=p_{1-}\ln (p_{2-}\hat y_0)\ .
\ee
The equations (\ref{eq2c1M2}), (\ref{eq2pcM2}) and (\ref{M3}) 
are compatible with each other (Note that $q_\pm$ is given in 
terms of $p_{1\pm}$ by (\ref{qq}). Let $\hat y_0$ and $q_+$ 
(or $p_{1+}$) be independent parameters. Then Eq.(\ref{eq2c1M2}) 
can be solved with respect to $q_-$:
\be
\label{M4}
q_-=q_-\left(q_+,\hat y_0\right)
\equiv { 1 \over {1 \over q_+} + {16\pi G\cdot 16\hat b' 
\over 3\sqrt{\hat y_0}}}\ .
\ee
Then  putting $\phi(\hat y_0)=p_{1+}\ln (p_{2+}\hat y_0)$ in 
(\ref{eq2pcM2}), we can solve the equation with respect to 
$p_{2+}$
\be
\label{M5}
p_{2+}=p_{2+}\left(q_+,\hat y_0\right)
\equiv {1 \over y_0}\e^{{1 \over 6\hat C p_1(q_1)} \cdot 
{\hat y_0^{3 \over 2} \over 8\pi G}\left(q_+ 
p_{1+}\left(q_+\right) - q_-\left(q_+,\hat y_0\right)
 p_{1-}\left(q_-\left(q_+,\hat y_0\right)\right)\right)}\ .
\ee
Here $p_{1\pm}\left(q_\pm\right)$ is defined by solving 
(\ref{qq}):
\be
\label{M6}
p_{1\pm}\left(q_\pm\right)=\sqrt{{3 \over 2}
 - {2k \over q_\pm^2}}\ .
\ee
Finally (\ref{M3}) can be solved with respect to $p_{2-}$:
\be
\label{M7}
p_{2-}={\left(p_{2+}\left(q_+,\hat y_0\right)\hat y_0
\right)^{p_1(q_1) \over 
p_{1-}\left(q_-\left(q_+,\hat y_0\right)\right)} \over y_0}\ .
\ee
When $k>0$, Eq.(\ref{M6}) gives a constraint
\be
\label{M8}
q_\pm^2 > {4k \over 3}\ .
\ee
As long as the constraint in (\ref{M8}) holds,  iterating the 
above procedure, we can obtain curved multi-brane  solutions.
Hence, we outlined the way to generalize two brane-world for multi-brane 
case.

\section{Discussion} 

In summary, we presented the generalization of quantum dilatonic 
brane-world\cite{NOO} where brane is flat, spherical (de Sitter) or 
hyperbolic and it is induced by quantum effects of CFT living on the brane.
In this generalization one may have two brane-worlds or even 
multi-brane-worlds which proves general character of scenario suggested in 
refs.\cite{HHR,NOZ} where instead of arbitrary brane tension added by 
hands the effective brane tension is produced by boundary quantum fields. 
What is more interesting the bulk solutions have analytical form,
 at least, for specific 
choice of bulk potential under consideration.

In classical dilatonic gravity the variety of brane-world solutions has 
been presented in ref.\cite{CEGH} where also the question of 
singularities has been discussed. The fine-tuned example of bulk
potential where one gets bulk solution which is not singular has been 
presented.
Let us consider if our solutions contain the curvature singularity or not. 
Multiplying $g_{(5)}^{\mu\nu}$ with the Einstein equation 
in the bulk:
\be
\label{Cur1}
0=R_{(5)\mu\nu}-{1 \over 2}\nabla_\mu\phi\nabla_\nu\phi 
 -{1 \over 2}g_{(5)\mu\nu}\left(R_{(5)} 
 - {1 \over 2}\nabla_\rho\phi\nabla^\rho\phi 
 + {12 \over l^2} + \Phi(\phi)\right)\ ,
\ee
which is obtained from $\SEH$ in (\ref{Stotal}), one gets
\be
\label{Cur2}
R_{(5)} = {1 \over 2}\nabla_\rho\phi\nabla^\rho\phi 
 - {5 \over 3}\left( {12 \over l^2} + \Phi(\phi)\right)\ .
\ee
Substituting expressions (\ref{assmp1}) 
and (\ref{assmp2}) into (\ref{Cur2}), we find
\be
\label{Cur3}
R_{(5)} = {p_1^2 \over 2y^2 f} - {5 \over 3}
\left(c_1(p_2y)^{ap_1} + c_2(p_2y)^{2ap_1}\right)\ .
\ee
Then for cases 1 $\sim$ 3, the scalar curvature $R_{(5)}$ is given 
by
\bea
\label{Curc1}
&\mbox{case 1}& \quad R_{(5)} 
= - {3 \over 2}{p_1^2 q^2 \over y} \\
\label{Curc2}
&\mbox{case 2}& \quad R_{(5)} 
= {8k \over y} - {2\tilde c_2 \over 3y^2} \\
\label{Curc3}
&\mbox{case 3}& \quad R_{(5)} 
= - {33\tilde c_1 \over 21\sqrt{y}} - {4k \over y}\ .
\eea
In all cases the singularity appears at $y=0$. 

In case 1, when $y\sim 0$ and the coordinates besides $y$ are 
fixed, the infinitesimally small distance $ds$ is given by
\be
\label{Cur4}
ds=\sqrt{f}dy \sim {dy \over q\sqrt{y}}\ ,
\ee
which tells that the distance between the brane and the 
singularity is finite. Then in cases of $k=0$ and $k<0$, 
the singularity is naked when we Wick re-rotate spacetime to 
Lorentzian signature. When $k>0$, the singularity is not 
exactly naked after the Wick re-rotation since the horizon 
is given by $y=0$, i.e. the horizon coincides with the 
curvature singularity. 

In case 2, the situation is not changed for $k=0$, $k>0$ 
 and $k<0$ with $\tilde c_2>0$ from that in case 1 and 
the distance between the brane and the singularity is finite 
since $ds\sim {dy \over \sqrt{y}}
\sqrt{3 \over 2\tilde c_2}$ when $y$ is small. 
When   $k<0$ with $\tilde c_2<0$, however, 
the singularity is not naked since there is a kind of horizon 
at $y={\tilde c_2 \over 2k}$, where ${1 \over f(y)}=0$. 
We should note the scalar curvature $R_{(5)}$ in (\ref{Curc2}) 
is finite. This tells that $y$ is not proper coordinate when 
$y\sim{\tilde c_2 \over 2k}$. If  new coordinate $\eta$ is introduced 
\be
\label{eta}
\eta^2\equiv 2\left(y - {\tilde c_2 \over 2k}\right)\ ,
\ee
the metric in (\ref{DP1}) is rewritten as follows,
\be
\label{DP1b}
ds^2=-{3 \over 4k}d\eta^2 + \left({\tilde c_2 \over 2k}
+ {\eta^2 \over 2}\right)
\sum_{i,j=1}^4\hat g_{ij}(x^k)dx^i dx^j\ . 
\ee
The radius of 4d manifold with negative $k$, whose metric is 
given by $\hat g_{ij}$, has a minimum ${\tilde c_2 \over 2k}$ 
at $\eta=0$, which 
corresponds to $y={\tilde c_2 \over 2k}$. The radius increases 
when $|\eta|$ increases. Therefore the spacetime can be regarded 
as a kind of wormhole, where two universes corresponding to 
$\eta>0$ and $\eta<0$, respectively, are joined at $\eta=0$. 

In case 3, the singularity is naked (the singularity 
is not exactly naked when $k>0$ as in case 1) in general 
and the distance between the brane and the horizon is finite 
except $k>0$ and $\tilde c_1<0$ case since there is a horizon 
at $\sqrt{y}=-{7k \over \tilde c_1}$ where the scalar curvature 
(\ref{Curc3}) is finite.
 
The price for having analytical bulk results (exactly solvable bulk 
potential) is the presence of (naked) singularity. One can, of course,
present the fine-tuned examples of bulk potential as in refs.\cite{NOO,CEGH}
where the problem of singularity does not appear. Moreover, bulk quantum 
effects may significally modify classical bulk configurations \cite{NOZ,NOZ2,
GPT} which presumbly may help in resolution of (naked) singularity problem.
 However, in such situation there are no analytical bulk solutions in
dilatonic gravity.

There are various ways to extend the results of present work.
First of all, one can construct multi-brane dilatonic solutions within 
the current scenario for another classes of bulk potential. However, this 
requires the application of numerical methods.
Second, it would be interesting to describe the details of brane-world 
anomaly driven inflation (with non-trivial dilaton) at late times when 
it should decay to standard FRW cosmology.
Third, within similar scenario 
one can consider dilatonic brane-world black holes which  
are currently under investigation.

\

\noindent{\bf Acknowledgements}

The work of S.D.O. has been supported in part by CONACyT (CP, 
ref.990356 and grant 28454E) and in part 
by RFBR and that of K.E.O. by RFBR.

\end{document}